# An objective criterion for evaluating new physical theories

Yefim Bakman, bakmanyef@gmail.com

August 14, 2025

Abstract: Currently, the value of a new physical theory is determined by whether it has been published in a prestigious journal. Yet, to publish an article in a prestigious journal, one must undergo a rigorous process of independent subjective peer review, with the success of the theory depending on the subjective opinion of reviewers who have proven themselves in developing old theories of physics. As a result, a closed system (gatekeeping) has been created that does not allow the penetration of new theories, especially those that call into question the established paradigm.

A subjective approach is not uniquely possible. Here, we propose an objective criterion for evaluating new physical theories and illustrate this criterion through an example.

## 1. Introduction

To construct a criterion for evaluating new physical theories, we use four indicators, as presented in Table 1.

Table 1.

| Number | Indicator | Description |
|--------|-----------|-------------|
| 1 | Entities | The number of postulates, laws, principles, and unique concepts (such as the fifth dimension, dark photons, strings, etc.) used in the work. |
| 2 | Solved problems | The number of problems that the theory explains by concrete physical mechanisms, serving as an indicator of explanatory power. |

| 3 | New problems | The number of new contradictions or questions created by the theory under discussion. |
|---|---|---|
| 4 | Integrity | The number of clear cause–effect relations between elements of the theory, representing the opposite of fragmentation. |

There is a broad consensus that Occam's razor should be included as a criterion for evaluating new theories in physics. Hence, the first indicator is the number of entities used in a theory. This term refers to postulates, laws, principles, and new concepts used in the work, such as the fifth dimension, dark photons, strings, etc. A reduction in the number of entities involved signals that a given theory is approaching the truth.

The above four indicators provide a simple report of a theory. A graphic diagram of the logical connections between the elements of the theory will be more meaningful and visual, as shown as the example in the next section.

## 2. An example of applying the proposed criterion

As an example, we use a new theory denoted as "A New Physical Paradigm," published in 2020 [1] [2]. This paradigm touches on various topics, but we will limit ourselves here to the basic questions of gravity.

Fig. 1 shows a logical diagram of the new paradigm, which is based on a fundamental medium (in Tesla's words, a "primary substance" [3]). The substance with a non-uniform density is a gravitational field.

In this paradigm, elementary particles are vortices of the same primary substance. Thus, gravity and quantum are closely related in the new paradigm.

From the four entities follow solutions to four problems: gravitational acceleration, dark matter, dark energy, and action at a distance.

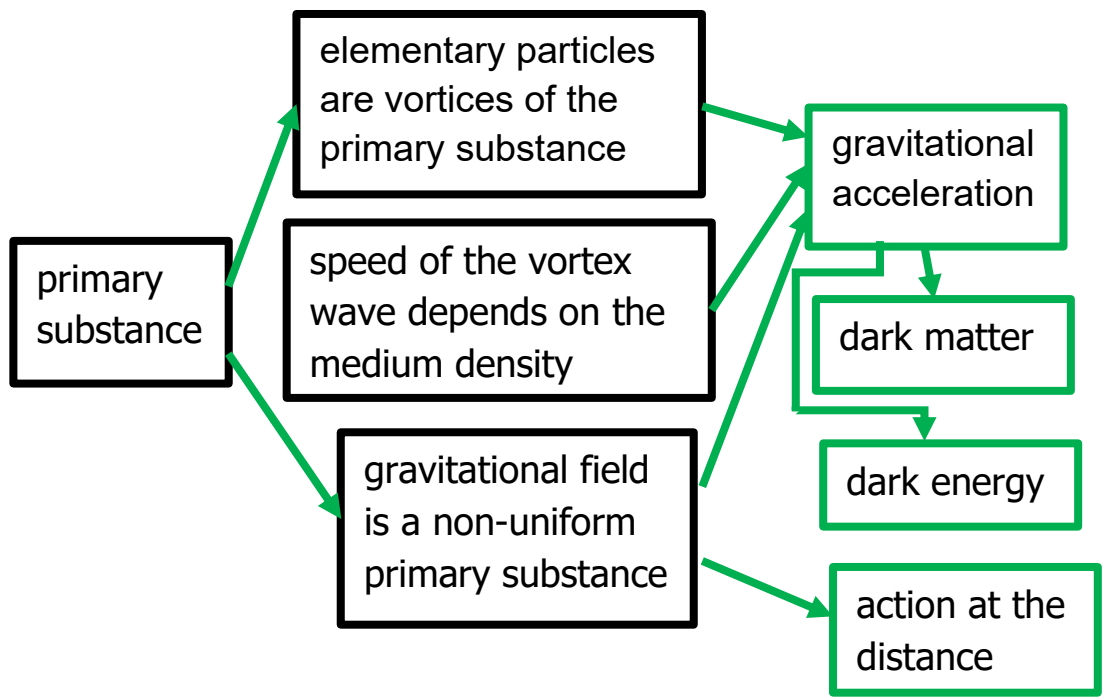

Fig. 1. Logical scheme of the new physical paradigm in relation to the topic of gravity. Entities (four) are shown in black frames, solved problems (four) are indicated in green frames, and cause–effect relations (eight) are presented by green arrows. New problems (zero in this case) would be indicated in red frames.

In addition to the universal medium and vortex particles, the new paradigm is based on the following insight:

The speed of the vortex wave depends on the density of the universal medium.

As a consequence, in a heterogeneous medium, vortex particles accelerate toward a higher density of the medium without the participation of contact forces, driven only by the difference in the velocities of the vortex wave [4].

The Earth's gravitational field is a consequence of the planet's atoms and molecules losing some of their mass or completely disintegrating into primary substance. Thus, a configuration is constantly being recreated in which the density of primary substance near the Earth's surface is higher than at a distance.

Table 2 shows scores for the four indicators of the new paradigm in relation to the topic of gravitation.

Table 2. Scores of objective indicators for the new paradigm with respect to gravitation.

| Indicator | Score |
|---|---|
| entities | 4 |
| resolved problems | 4 |
| new problems | 0 |
| integrity | 8 |

For comparison, the dominant paradigm incorporates numerous entities to explain the structure of hadrons: six types of quarks and eight types of gluons to hold quarks together. Yet, to explain the phenomena of gravity and dark matter/energy, it is not possible to select entities, despite the freedom of their choice.

## 3. Conclusion

A graphical representation depicting the structure of a theory can show relationships between entities, problems, and mechanisms. We propose that the scientific community adopt this format as a standard supplement to theoretical articles in order to provide editors and readers with a visual representation of the theory structure.